\documentclass{emulateapj}
\begin{document}
\voffset=-0.25in

\title{Three-dimensional Evolution of Solar Wind 
during Solar Cycles 22--24}

\author{P.K. MANOHARAN}
\affil{Radio Astronomy Centre \\
National Centre for Radio Astrophysics \\
Tata Institute of Fundamental Research\\
Udhagamandalam (Ooty) 643 001, India}
\email{mano@ncra.tifr.res.in}

\shorttitle{3-D Changes in Solar Wind - Solar Cycles 22 to 24}
\shortauthors{P.K. Manoharan}

\begin{abstract}

This paper presents the analysis of three-dimensional evolution of solar
wind density turbulence and speed at various levels of solar activity 
between solar cycles 22 and 24. The solar wind data used in this study 
has been obtained from interplanetary scintillation (IPS) measurements 
made at the Ooty Radio Telescope, operating at 327 MHz. Results show that
(i) on the average, there was a downward trend in density turbulence from 
the maximum  of cycle 22 to the deep minimum phase of cycle 23; (2) the 
scattering diameter of the corona around the Sun shrunk steadily 
towards the Sun, starting from 2003 to the smallest size at the deepest 
minimum, and it corresponded to a reduction of $\sim$50\% in density 
turbulence between maximum and minimum phases of cycle 23; (3) The 
latitudinal distribution of solar wind speed was significantly different 
between minima of cycles 22 and 23. At the minimum phase of solar cycle
22, when the underlying solar magnetic field was simple and nearly dipole
in nature, the high-speed streams were observed from poles to $\sim$30$^\circ$ 
latitudes in both hemispheres. In contrast, in the long-decay phase of cycle 23, 
the sources of high-speed wind at both poles, in accordance with the weak polar 
fields, occupied narrow latitude belts from poles to $\sim$60$^\circ$ latitudes. 
Moreover, in agreement with the large amplitude of 
heliospheric current sheet, the low-speed wind prevailed the low- and mid-latitude
regions of the heliosphere. (4) At the transition phase between cycles 23 and 
24, the high levels of density and density turbulence were observed close to 
the heliospheric equator and the low-speed speed wind extended from equatorial- 
to mid-latitude regions. The above results in comparison with Ulysses and other
in-situ measurements suggest that the source of solar wind has changed globally, 
with the important implication that the supply of mass and energy from the Sun 
to the interplanetary space has significantly reduced in the prolonged period
of low level of solar activity. The IPS results are consistent with the onset 
and growth of the current solar cycle 24, starting from middle of 2009. However, 
the width of the high-speed wind at the northern high latitudes has almost
disappeared and indicates that the ascending phase of the current cycle has
almost reached near to the maximum phase at the northern hemisphere of the Sun. 
But, in the southern part of the hemisphere, the solar activity has yet to 
develop and/or increase.

\end{abstract}


\section{Introduction}

The behavior of the Sun before the transition between solar cycles 23 and 
24 exhibited very unusual longest and deepest minimum of activity. At the 
long-decay phase, the magnetic configuration of the Sun went through 
remarkable changes, which were different from those observed at the 
corresponding previous minimum phase. For example, the number of days
without sunspot was large (more than 800 days at the deep minimum phase, 
compared to $\sim$300 days at the minimum of cycle 22) and the fluxes of
extreme ultraviolet, soft X-ray, and radio intensity at 10.7 cm reached 
the lowest levels. Moreover, unusually long-lasting high-speed streams from 
low-latitude coronal holes of open magnetic field lines and their interaction 
with low-speed flows from closed field corona resulted in a complex heliosphere
of low speed and density. The reasons for the long duration as well as depth of 
the peculiar minimum have been much studied, in terms of solar 
interior characteristics, manifestation of sunspot, polar field strength, 
transmission of solar wind, shaping of heliosphere, geo-effectiveness, 
solar irradiance, etc. \citep[see, e.g.,][]{basu2010, feynman2011, 
jian2011, lallement2010, lee2009, lo2010, mano2010b, mccomas2008, 
smith2008, tapping2011, tokumaru2010}.
An international campaign on `Whole Heliosphere Interval' was also
organized to study the three-dimensional aspects of 
`solar-heliospheric-planetary connected system'. The period of the
campaign covered a part of the deep minimum phase (i.e.,
Carrington Rotation 2068, 20 March -- 16 April 2008) and included
studies from low solar photosphere, through interplanetary space, 
and down to Earth's mesosphere (e.g., Thompson et al. 2011).

The extended decay of the cycle 23 and the later than expected onset 
of cycle 24, provided opportunity to examine the link between the 
solar activity and three-dimensional heliosphere. In principle, the 
three-dimensional view of the solar wind can be inferred by the radio 
remote sensing interplanetary scintillation (IPS) technique, which can 
provide estimates of solar wind speed and density turbulence. Several 
authors employed the IPS technique and studied the changes of solar wind
as functions of helio latitude and distance as well as over solar cycle and 
compared them with three-dimensional coronal density and magnetic field 
structures on the Sun (e.g., Rickett \& Coles 1991; Kojima \& Kakinuma 1990; 
Manoharan 1993, 1995, 1997; Manoharan et al. 1994). 
Recently, based on IPS observations from STELab, a distinct decrease 
in solar wind speed at the high-latitude regions of the heliosphere 
was revealed for the solar cycle 23 \citep{tokumaru2010}. 

In the present study, the large IPS data base collected from the Ooty 
Radio Telescope, operated by Radio Astronomy Centre, Tata Institute of 
Fundamental Research, India, at 327 MHz (Swarup et al. 1971), has been 
employed to analyze the three-dimensional distributions of solar wind 
density turbulence and speed at different levels of solar activity 
during 1985--2011. It is the detailed study of preliminary results 
presented in a conference proceeding \citep{mano2010b} and covers more
than 2 solar cycles. The careful examination of consequences of various 
levels of Sun's activity embedded in the solar wind gives the insight 
into the fundamental physical processes involved in shaping the 
heliosphere.  The paper is structured as follows: a brief description 
of IPS observation is given in Section 2. The next section covers 
the results on distribution of density turbulence at different 
phases of solar cycles 22--24. The latitudinal changes of solar
wind speed from IPS and Ulysses spacecraft are discussed in Section 
4. In Section 5, measurements obtained from near-Earth spacecraft 
are discussed. Section 6 summaries the results and discussion.

\section{Interplanetary Scintillation}

The IPS technique exploits the scattering of radio waves from a distant 
compact source of angular size, $\Theta \leq 500$ milliarcsec (e.g., a 
radio galaxy or quasar), by the density turbulence in the solar wind 
\citep[e.g.,][]{hewish1964,coles1978,mano1990,kojima2004}. 
The measurable quantity in an IPS observation is the time series of 
intensity fluctuations, resulting from the radio-wave scattering 
caused by the plasma density irregularities convected at the speed 
of the solar wind and it includes irregularity structure of spatial 
scales from about the size of Fresnel scale and down below. For 
example, the IPS used in this study, at 327 MHz, can probe turbulence 
scales below 400 km.  
A suitable calibration of the temporal spectrum of 
intensity fluctuations can provide the solar wind speed
\citep[e.g.,][]{mano1990, mano2000, tokumaru1994, tokumaru2011, yamauchi1996, 
yamauchi1998, liu2010} 
and the scintillation index, {\it m} 
\citep[e.g.,][]{coles1978, mano1993, asai1998}. 
The scintillation index, $m$, is a measure of electron-density 
turbulence in the solar wind ($m^2 \sim \int\delta N^{2}_{e}(z)~dz$), 
along the line of sight ({\it z}) to the radio source (e.g., 
Manoharan 1993; Manoharan et al. 2000). 

An IPS measurement therefore represents the integration along the 
line of sight to the radio source. However, since most of the 
scattering power is concentrated at the point of closest approach 
of the line of sight to the Sun, as given by the typical steep radial fall 
of density turbulence, $\delta N^2_e(R)$ $\sim$ $R^{-4}$, the IPS 
basically probes properties of the solar wind at the region of the 
closest solar offset on the source--Earth path
\citep{coles1978,mano1993}. In fact,
the line of sight integration can pose a problem when a short-lived
solar wind transient of enhanced density turbulence and speed with
respect to the ambient solar wind is studied based on a single IPS 
observation, which may lead to positional uncertainty along the 
integration path. However in the present study, the high-level of
scintillation associated with solar wind transients, e.g., coronal
mass ejections (CMEs), have been removed. In the case of 
slowly varying large-scale ambient solar wind, an IPS observation 
can only systematically underestimate the solar wind by $\sim$5--10\% 
\citep{mano1994,mano2000,yamauchi1996}.
Moreover, when a day-to-day monitoring of IPS is made on a large 
number of scintillators having different lines of sight, cutting 
across different parts of the heliosphere, it can probe the 
substantial portion of the inner heliosphere. Such data sets are 
extremely useful to study the evolution of large-scale features 
of the solar wind over an extended period of time 
\citep{mano2006, mano2010a, mano2010b}.
Since the primary aim of the present study is to understand the 
three-dimensional changes of large-scale structure of solar wind 
over a long period of time (e.g., at time scales of fraction of a 
year to solar cycle), no attempt has been made to remove the systematic
line-of-sight integration effect included in the IPS observation.

In this study, a large amount of IPS data collected during 1985--2011 
from the Ooty Radio Telescope (ORT), operating at 327 MHz 
\citep{swarup1971}, has been employed.  This set of IPS data 
from Ooty probed the solar wind in the heliocentric distance range 
of $\sim$10--250 solar radii ($R_\odot$, 1 $R_\odot$ = $6.96 \times
10^5$~km, 1~AU~$\approx 215 R_\odot$) and at all heliographic 
latitudes. It allowed the study of the three-dimensional 
evolution of the heliosphere over solar cycles 22 to 24. 
It may be noted that before the upgrade of the feed system
of the ORT around middle of 1992, everyday IPS measurements 
were limited to a small number of radio sources ($\sim$40--50 
radio sources). The increased sensitivity of the upgraded system 
however enabled the observations of $\sim$300 sources per day and 
this source count steadily increased to the present monitoring of a grid 
of more than 1000 sources per day. Currently, the ORT is being 
upgraded and the number of IPS sources observed per day is expected to 
increase by $\sim$4 to 5 fold by the end of this year 
(Prasad \& Subrahmanya 2011).
The regular monitoring of IPS so far made at Ooty covers an extended 
period of more than 2 solar cycles and it has led to the detection of
more than 3500 scintillating sources at 327 MHz, over the entire right 
ascension range. Results on compact component structure of these 
sources are in preparation \citep{mano2012}. 

\begin{figure*}
\begin{center}
\includegraphics[width=12.0cm]{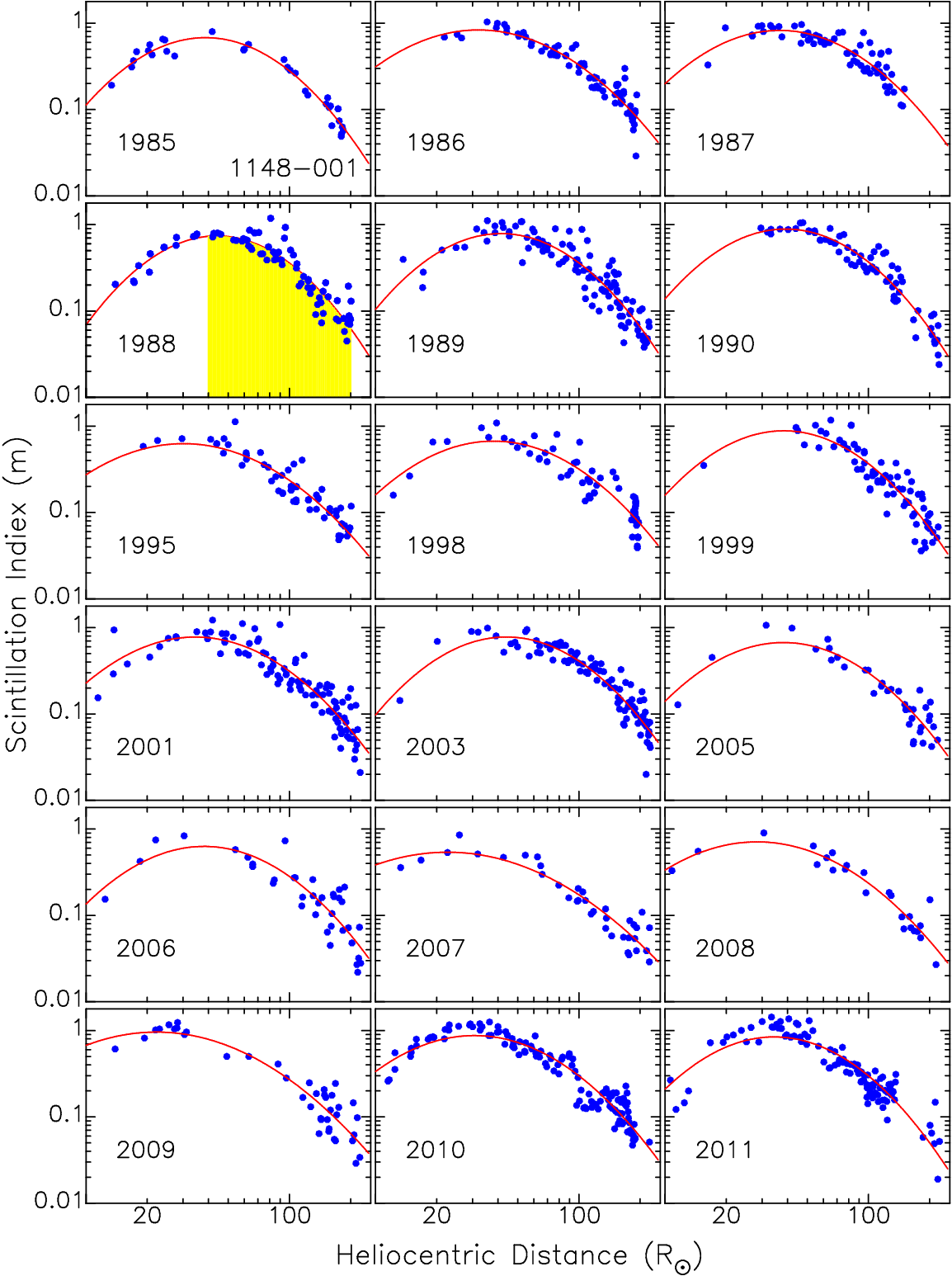}
\caption{Scintillation index as a function of distance from the Sun for
the radio quasar 1148-001 measured from the Ooty Radio Telescope, 
operating at 327 MHz, over the years 1985--2011. The peak of $m-R$ curve, 
close to unity, shows that the radio quasar 1148-001 is compact, which in 
fact has an equivalent diameter of $\sim$15 milliarcsec. In each plot, 
the best-fit to the data points is shown by a continuous curve. The shaded
portion shown on the plot of year 1988 indicates the radial distance range 
over which the area has been computed as given in equation 1.}
\end{center}
\end{figure*}

\section{Solar Cycle Changes of Solar Wind Density}

\subsection{Scintillation Index and Large-scale Density Turbulence}

The degree of interplanetary scintillation is given by the scintillation 
index, {\it m = rms of intensity fluctuations/mean intensity of the source},
which is a measure of density turbulence in the solar wind,
($m^2(R) \sim \int \delta N^{2}_{e}(R)~dR$).
Figure 1 shows the scintillation index measurements made with the Ooty 
Radio Telescope at 327 MHz in the heliocentric distance range of 10--250 
$R_\odot$ on a compact radio quasar 1148-001, for selected years between 
1985 and 2011. The above radio quasar has a compact component of size 
$\sim$15 milliarcsec \citep{mano1995a}. 
As indicated by the peak value of the scintillation index close to 
unity, the compact component of the above source contains more than 
90\% of its total flux density. It is to be noted that for a given 
radio source, the entire scintillation index profile, $m(R)$, is 
attenuated by the brightness-distribution function associated with 
the size of the scintillating component. Therefore, the peak value 
of scintillation can vary between unity and null, respectively, for 
an ideal point source and a non-scintillator 
\citep{coles1978, mano1994, mano2006}. 
Since in the above plots, the enhanced 
scintillations caused by intense solar wind transient events (e.g., 
coronal mass ejections, refer to Manoharan et al. 2000) have been 
excluded, these plots represent the average condition of the ambient density 
turbulence of the heliosphere at different phases of solar cycles 
22--24. Moreover, the radio quasar 1148-001 is an ecliptic source 
and its IPS measurements are confined to the equatorial region of 
the heliosphere. Therefore, Figure 1 represents solar wind changes 
occurred in the equatorial belt of the heliosphere.

\begin{figure}
\begin{center}
\includegraphics[width=6.0cm,angle=-90]{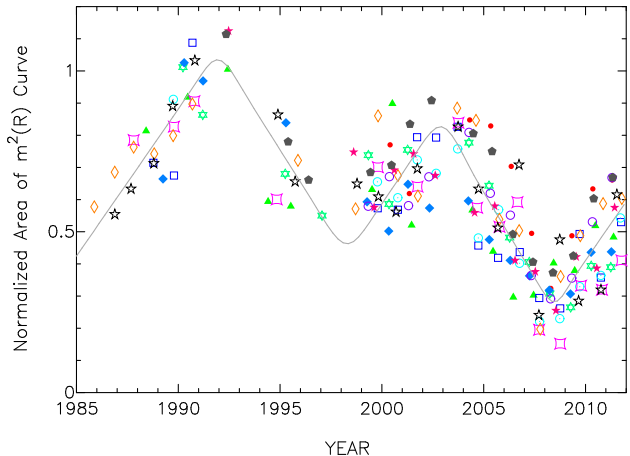}
\caption{The area under $m^2(R)$ profile (equation 1) plotted as a
function of year. The area has been computed in the weak-scattering
region in the distance range of 40--200 $R_\odot$, as indicated by the 
shaded area shown in Figure 1 (refer to 1988 plot). Different symbols correspond 
to different scintillating sources. These are strongly scintillating 
sources, having equivalent diameter $\leq$100 milliarcsec and their 
compact components contain more than 50\% of the source total flux density.
The continuous line is the segment-wise best fit to the data points.
It is clear from the plot that the deepest minimum of of the solar
cycle 23 is revealed by the lowest level of density turbulence around 
middle of 2008.}
\end{center}
\end{figure}


In Figure 1, the best-fit to the data points is shown by a continuous
curve. As shown in the figure, the level of scintillation as the Sun
is approached, increases to a peak value near a distance of 
$R \approx 40~R_\odot$, and then decreases for further closer solar 
offsets, where the scattering becomes strong and saturated (e.g., 
Manoharan 1993). The turnover point of the scintillation index moves 
close to the Sun as the observing frequency is increased. However for 
a given observing frequency, it depends only on the level of turbulence. 
In the above log-log `$m$-$R$' plots shown in Figure 1, the peak value of
scintillation index increases or decreases, respectively, in tune with 
maximum or minimum of solar activity. However, in order to infer the 
global characteristics of density turbulence in the heliosphere, the 
area under the best fit $m^2(R)$ profile has been computed for several 
scintillators as given by,
\begin{eqnarray}
A = \int_{R=40 R_\odot}^{R=200 R_\odot} m^2(R)~dR.
\end{eqnarray}

For each source, the area has been computed in the weak-scattering 
regime ($R$ in the range $\approx 40-200 R_\odot$). Since most of the 
scattering power is contained within $\sim200~R_\odot$, the above
summation provides an average quantitative estimate of the large-scale 
density fluctuations in the inner heliosphere of radius $\sim$0.2--1 AU
over a period of about 6 months. 
Figure 2 shows the area of $m^2(R)$ curve plotted against the year 
for 12 compact scintillating sources. These sources have been 
selected on the criteria: (i) strong scintillator having the compact 
component of size $\Theta \leq 100$ milliarcsec, (ii) more than 50\% of 
the total flux density of the source lies in the compact component 
(or the peak of the scintillation index curve is $\gtrsim$0.5), and
(iii) observations of these sources fall in the ecliptic plane 
(i.e., in the equatorial region of the heliosphere). For each 
source, the computed area has been normalized by its average fraction 
of scintillating flux density (i.e., the ratio between the average of
scintillating component flux obtained from all the available observations
over the period between 1985 and 2011 and the total flux density of 
the source), which is a value between unity and 0.5, respectively, 
corresponding to an ideal compact source and a marginally broad 
source. In Figure 2, the data points in a year show scatter along 
the vertical axis and it is due to the changes in large-scale solar 
activity from one source to the other. Nevertheless, a smooth curve has
been obtained by making segment-wise least-square fit to the data 
points (refer to thin line in Figure 2) and the solar cycle changes 
are evident in the plot. The large-scale density turbulence gradually 
increases from the minimum phase of the solar cycle 22, around 
1986, and peaks near the consecutive maximum around the year 1991. A 
reduced plasma turbulence of $\sim$70\% level is observed at the minima of both 
cycles 21 and 22, respectively, around 1986 and 1996. However, the 
density turbulence levels at the maximum and minimum of the cycle 23, 
respectively around 2002 and 2008, are observed to be lower than the 
corresponding values observed in the cycle 22. A decreasing trend in 
the turbulence level is observed from the year 1990 to 2009. At the
transition between cycles 23 and 24 (i.e., between years 2008 and 2009),
the density turbulence attains the lowest value of the solar cycle 23.  
A significant broad peak seen in density 
turbulence around the year 2003 is likely to be associated with the 
co-rotating interaction regions (also refer to Figure 4), which are
caused by the persistent mid-latitude coronal 
holes on the Sun (Manoharan 2010b; Abramenako et al. 2010)

\subsection{Radial Dependence of Density Turbulence}

In the weak-scattering region at $R > 40~R_\odot$ (refer to Figure 1), 
the gradual decrease in the scintillation with distance from the Sun 
is linearly related to the typical radial fall of density turbulence,
$m^2 \sim \int\delta N^{2}_{e}(R)~dR \sim \int C^{2}_{N}(R)~dR$, and 
from several scattering observations in the inner heliosphere, it is 
shown to follow the form, $\delta N^{2}_{e}(R) \sim R^{-4}$ 
\citep[e.g.,][]{mano1993, coles1995, coles1996}. Thus, the quantity 
$C^{2}_{N}(R)$ is the measure of mean-square relative fluctuation 
in plasma density, $N_e(R)$, and it scales the spatial spectrum of 
density turbulence, $\Phi(q,R) = C^{2}_{N}(R)~q^{-p}$, where $q$ is
the three-dimensional wavenumber. In the case of undisturbed solar 
wind, the spectrum takes the form, $\Phi(q,R) \sim q^{-3.3}$ and it
gets attenuated at the high-wavenumber portion by the dissipative-scale 
(or inner-scale) size increasing linearly with heliocentric distance, 
$S_i \approx (R/R_\odot)^{1.0\pm0.1}$ km at $R \leq 100 R_\odot$ 
\citep{mano1988, coles1989, mano1994, yamauchi1998}. However, at 
larger distances from the Sun, $\sim$100--200~$R_\odot$, the inner 
scale tends to stay at a constant value of $S_i \approx 100$ km 
\citep{mano1988, mano1994}. The inner-scale cutoff is attributed to 
occur near the ion (proton) inertial scale, which is determined by 
the local Alfven speed and ion cyclotron frequency 
\citep[e.g.,][]{coles1989, coles1991, mano1994}. 
Therefore, the contributions to the overall turbulence spectrum
include magnetic fluctuations (associated density structures) and 
density fluctuations and their radial changes as a function of 
scale size.

For example, using log-log plots shown in Figure 1, the radial 
dependence of scintillation can be obtained from the linear slope 
of $m-R$ curve at distances $>$40 $R_\odot$. 
Thus, the radial dependence of scintillation, 
$ m = m_0R^{-\alpha}$ ($m_0$ 
is the scintillation index at the unit distance from the Sun), 
corresponds to changes in density turbulence with heliocentric 
distance. The $m-R$ curves of compact sources used in Figure 2 
have been employed to estimate the radial fall of density 
turbulence. For each year, the weak-scattering portion of $m-R$ 
curve (at distances $\geq50~R_\odot$) has been fitted with a 
least-square straight-line fit. The magnitude of slopes ($\alpha$) 
obtained from all the sources vary between 1.7 and 2.1, with an 
average of $\sim$1.8.  In the period considered between 1985 and 
2011, the slopes do not show significant changes in correlation 
with the solar cycle. However, marginal changes are observed in 
the radial dependence of density turbulence between periods before 
2003 and later. In the prolonged minimum phase, after the year 2003, 
the radial fall shows an average indolent (or slow) slope of 
$\alpha\approx 1.7$ and a steeper slope (i.e., $\alpha\approx 1.9$) 
is observed for years before 2003, i.e., over the period of cycle 
22 and until about the maximum of cycle 23. 

As it was stated in Section 3.1, the integration along the 
line-of-sight would reduce the radial power by one unit 
(Readhead et al. 1978; Manoharan 1993).  Therefore, the 
`{\it distance-density turbulence}' relationship, 
$\delta N^2_e(R) \sim R^{-\beta}$, is related to the slope of the 
scintillation index curve by $2\alpha+1 = \beta $. The steep slope 
observed, $\alpha \approx 2.1$, indicates that the density turbulence 
falls off rapidly with distance, which is about 
$\delta N_{e}(R) \sim R^{-2.6}$. Whereas the value of shallow slope, 
$\alpha \approx 1.7$, suggests a typical $R^{-2.2}$ dependence, which 
is close to the symmetrically expanding solar wind. These results are 
in agreement with the earlier findings obtained for the solar cycle 21 
\citep{mano1993, coles1995, coles1996}. 
The marginally less rapid slope observed in the long-deep minimum
phase, $R^{-2.2}$ in comparison to the average steep slope of $R^{-2.3}$, 
indicates that the slow fall of solar wind is likely linked to the 
energy and mass fluxes associated with the solar wind originating at the 
base of the corona and its interaction with the background flow in the 
inner heliosphere. Thus, the excessive turbulence is possibly due to the 
interacting flows at $R > 100 R_\odot$, largely due to mid and low latitude 
coronal holes, which in fact persisted in the long-decay phase.

However, the mechanism of observed rapid fall in the range, $\delta N^2_{e}(R) 
\sim R^{-(\beta \approx 4~{\rm to}~5)}$, and its possible influence beyond 1 AU 
are required to be addressed. For example, a study of radial dependence of density 
turbulence by Bellamy et al.~(2005), based on Voyager 2 spacecraft data, has 
shown much lesser fall in the outer heliosphere. Below 30 AU, small slopes of
$\beta$ = 2.3$\pm$0.9 and $\beta$ = 3.3$\pm$0.9 are reported, respectively,
for 96-s and 192-s sampling intervals, which correspond to spatial scales of
about one to two orders of magnitude larger than scales probed by the IPS 
observation. The above study also reveals that the level of $C^{2}_{N}(R)$ 
drastically changes with the employed sampling rate (i.e., probed scale size
of turbulence). Thus, at distances beyond 1 AU, the radial dependence of
turbulence spectrum, $\Phi(q,R)$ (also $C^{2}_{N}(R)$), shows flattening, 
enhancement, and dissipation as function of spatial scales and tends to be 
independent of regular expansion of the background solar wind 
\citep{bellamy2005, hunana2008, zank1996}.

\begin{figure*}
\begin{center}
\includegraphics[width=12.0cm]{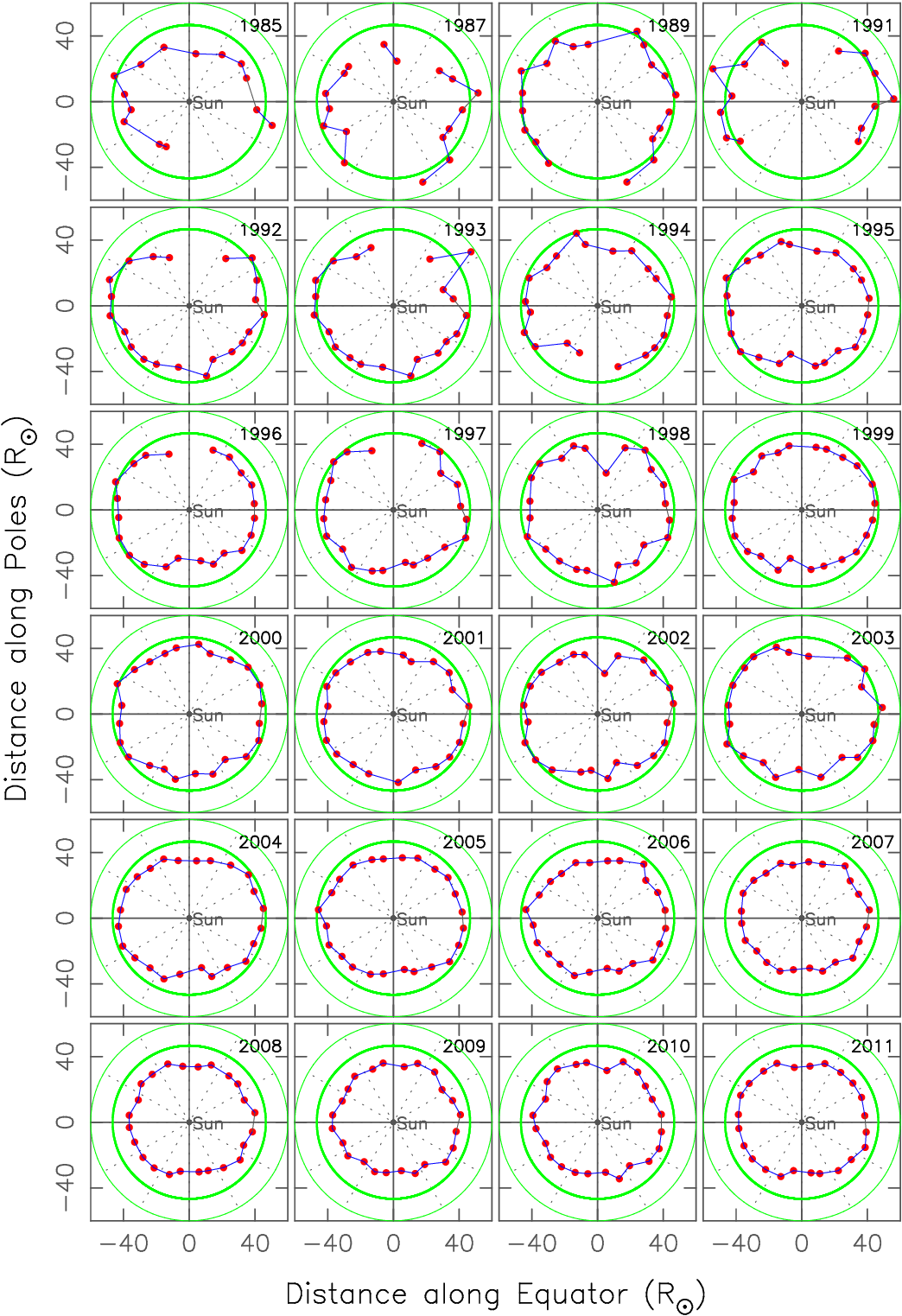}
\caption{Constant $\delta N_{e}(R)$ contours plotted on 
`north-east-south-west-north' polar diagram for selected years 
between 1985 and 2011. The last plot includes data up to November 
2011. In each plot, reference circles are drawn, respectively, 
at radii 45 and 60 $R_\odot$ and 30-degree latitude sectors are 
shown by dotted lines. During 2008-09, the scattering diameter of 
the corona has shrunken to the smallest size and the solar activity 
touched the deepest minimum of solar cycle 23. The increase in the 
diameter of the contour in the year 2011 evidently reveals the 
progression of cycle 24.}
\end{center}
\end{figure*}

\subsection{Scattering Diameter of the Corona}

The routine monitoring of IPS at Ooty, on a large number of compact 
radio sources during the period 1985--2011, has been used to estimate
constant density turbulence contours at different phases of solar 
cycles 22 to 24 \citep{mano2006, mano2010b}. 
At an observing frequency, the peak of the scintillation curve, {\it m--R}
(refer to Figure 1), depends on the fixed level of density turbulence and 
the heliographic coordinates of several peaks distributed around the Sun 
are useful to find the contour of the constant (or same) level of density 
turbulence (${\delta N_e(R)}$) in the heliosphere (e.g., Manoharan 1993; 
Coles 1978). At 327 MHz, the turnover of scintillation occurs around
$\sim$40$R_\odot$. In Figure 1, although these are log-log plots, a
careful examination reveals that the peak point of the scintillation
curve (i.e., constant value of ${\delta N_e(R)}$) moves close or away 
from the Sun, respectively, at the minimum
or maximum phase of the solar cycle. Therefore at 327 MHz, the 
estimation of $m-R$ curves of several ecliptic and out-of-ecliptic 
sources in a year and their corresponding peaks located at different 
latitudes (as well as heliocentric distances) can provide the trace of
constant level of density turbulence around the Sun in the distance
range of 30--50 $R_\odot$.

In a year for each source, the average peak of the scintillation and 
its heliographic coordinates are obtained and plotted around the Sun
on a north-east-south-west-north radial diagram. As shown in Figure 3,
the latitude-bin average of peaks at different latitudes and the line 
joining them can provide the constant density turbulence contour around 
the Sun (e.g., Manoharan 1993). Although this figure includes average 
constant ${\delta N_e(R)}$ contours for selected years between 1985 and 
2011, the data coverage was sparse during the solar cycle 22. It was
due to the fact that in the early days observations at Ooty less concentration was 
given to IPS below 50 $R_\odot$ and measurements were mostly limited to 
the weak-scattering regime, $> 50 R_\odot$. However, starting from the 
second half of cycle 
22, IPS measurements were carried out over the full distance range of 
10--250 $R_\odot$ and they enabled the determination of transition 
of scintillation for sources favorably positioned with respect 
to the Sun. In Figure 3, the mean 
traces of constant level of density turbulence indicate that in general at 
the polar regions, the constant level of turbulence is observed closer 
to the Sun than at the equatorial region. It is in agreement that at
the descending phase of the activity,
the low-density (i.e., high-speed) streams from coronal holes fill 
the polar regions, leading to low level of density turbulence and
constant ${\delta N_e(R)}$ contour moves close to the Sun to compensate 
the fall in scattering power at the high-latitude region of the corona.
However, as the Sun approaches towards the minimum of activity, 
the size of the coronal hole increases and extends to mid 
latitudes \citep{mano1993, coles1995, coles1996}. 
In such a phase, the density turbulence contour takes a shape close to 
an ellipse, in which the contour at the mid-latitude part also moves 
close to Sun (Figure 3, refer to 1994--96 plots). In the ascending phase, the 
ratio between poleward and equatorial diameters however gradually decreases 
from minimum to maximum phase of the cycle (e.g., Manoharan 1993; 
Coles et al. 1995) and the ratio of diameters tends to unity at the maximum 
of the cycle. The solar wind flow thus attains a spherically symmetrical 
flow, as revealed by the almost circularly symmetrical distribution
of streamers at the maximum phase of the corona observed during the total 
solar eclipse (e.g., Koutchmy et al. 1992).

\begin{figure*}
\begin{center}
\includegraphics[width=8.0cm,angle=-90.0]{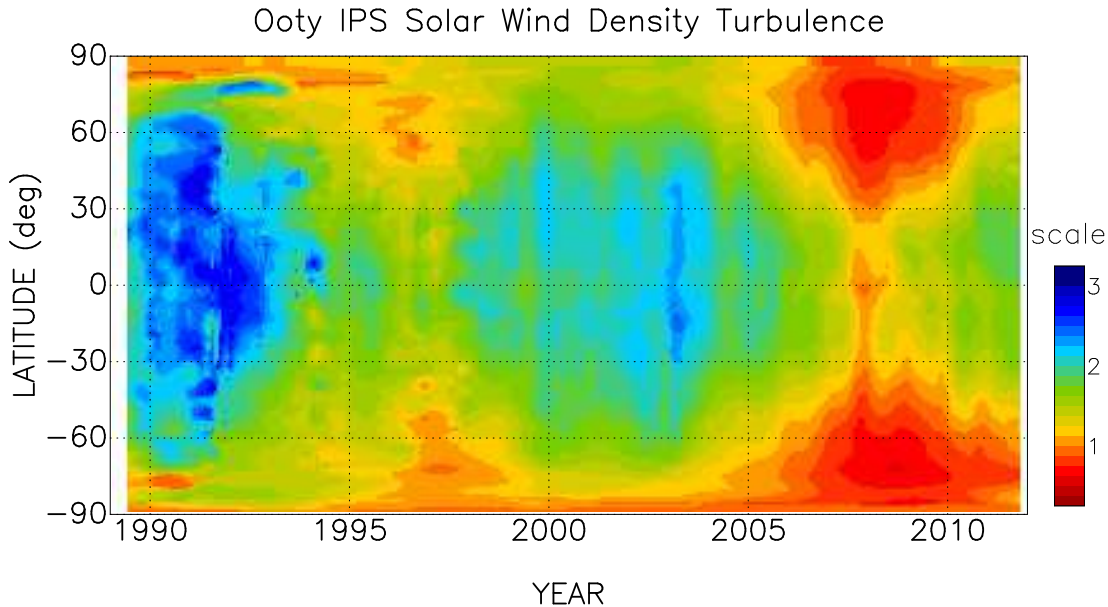}
\caption{Helio latitudinal distribution of solar wind density 
turbulence as a function of year. This image has been made by 
allowing data in the heliocentric distance range of 75--125 
R$_\odot$ and it shows the condition of the heliosphere at the midway 
between the Sun and Earth. The distribution of density turbulence
shows remarkable change between minima of solar cycles 22 and 23.}
\end{center}
\end{figure*}

The interesting result of this study is that after about the year 2003, 
the overall scattering diameter of the corona has gradually decreased
with respect to the center of the Sun and a given level of turbulence
has steadily moved close to the Sun at all latitudes. The diameter of 
contours during the 
years 2007--2009 appear symmetrical, but much smaller than the contours
observed in the cycle 22. The typical radial dependence of turbulence,
$C^{2}_{N_e}(R) \sim R^{-4}$ to $R^{-4.4}$, 
suggests that the scattering has
remained nearly same at low-latitude regions between 1995 and 2003, 
but, steadily decreased to $\sim$60\% level around middle of 2009, 
where the deepest minimum of cycle 23 was observed.  The shrinking of 
scattering diameter of the corona is in agreement with the measured 
level of scintillation displayed in Figure 2, which also shows
the overall gradual reduction in scattering power between the years 
1986 and 2010.  The equatorial diameter starts to increase between 
2010 and 2011, indicating the onset and progress of the solar cycle 24.
The steady and significant long-term trend of shrinking 
of scattering power of heliosphere means a natural reduction in supply 
of mass and energy at the low corona.

\subsection{Latitudinal Distribution of Density Turbulence }

At Ooty, IPS on a source is measured at different solar offsets.
For each source, the scintillation index plot ($m(R))$ in a year, 
as plotted in Figure 1, can be least-square fitted with an average curve. 
The ratio between the observed and expected (or fitted) indexes at 
a given heliocentric distance, 
{\it g} = {\it observed scintillation}/{\it average expected 
scintillation}, can be used to assess 
the level of density turbulence of the slowly varying solar 
wind structures in comparison with the ambient (or background)
solar wind plasma \citep{mano2006}.
The above normalized scintillation index, {\it g}, is independent 
of systematic fall of {\it m} with solar offset and source-size 
attenuation. 
For example, a value of the normalized index close to unity 
($g\approx 1$) represents the undisturbed (or background)
condition of the solar wind, $g > 1$ corresponds to the enhanced 
level of density turbulence, and, $g < 1$ indicates the reduction 
of turbulence in the solar wind. An IPS measurement is therefore sensitive to 
detect even a small fractional change in the density turbulence 
and the routine monitoring of IPS on a grid of sources can easily 
detect the large-scale structures over a given period of time
\citep{mano2006}.

For each source, the best fit to $m-R$ curve of a given year is 
estimated and it is used to eliminate the intense transients caused 
by the CMEs. The transients removed data sets of a given source, available 
for the period between 1985 and 2011, are then combined and the overall 
least-square fit is computed. Thus, the value of {\it g} obtained 
from the fitted curve and each observed scintillation index allows an 
easy comparison of levels of turbulence measured on a number of IPS 
sources at different time periods and it enables the detection of 
slowly varying large-scale density structures in the solar wind.
Figure 4 displays the `latitude--year' plot constructed from the
normalized scintillation index estimates obtained from the continues
Ooty IPS measurements between 1989 and later part of 2011 
\citep{mano2000, mano2001, mano2006}. 
In a broad sense, this  plot resembles the `{\it sunspot butterfly}' 
diagram and reveals the major large-scale structural changes of 
the heliosphere occurred through solar cycles 22 to 24. Since
the density turbulence ($\delta N^{2}_{e}$) and {\it g} are 
linearly related only in the weak-scattering region, which 
typically occurs at distances $>$40 $R_\odot$ for IPS at 327 MHz 
(e.g., Manoharan 1993), in the above image we have employed 
measurements made in the distance range of 75--125 $R_\odot$.
Therefore, the above latitudinal plot represents the density 
turbulence of the inner heliosphere at a radius of about midway 
between the Sun and the Earth.

\begin{figure*}
\begin{center}
\includegraphics[width=8.7cm,angle=-90]{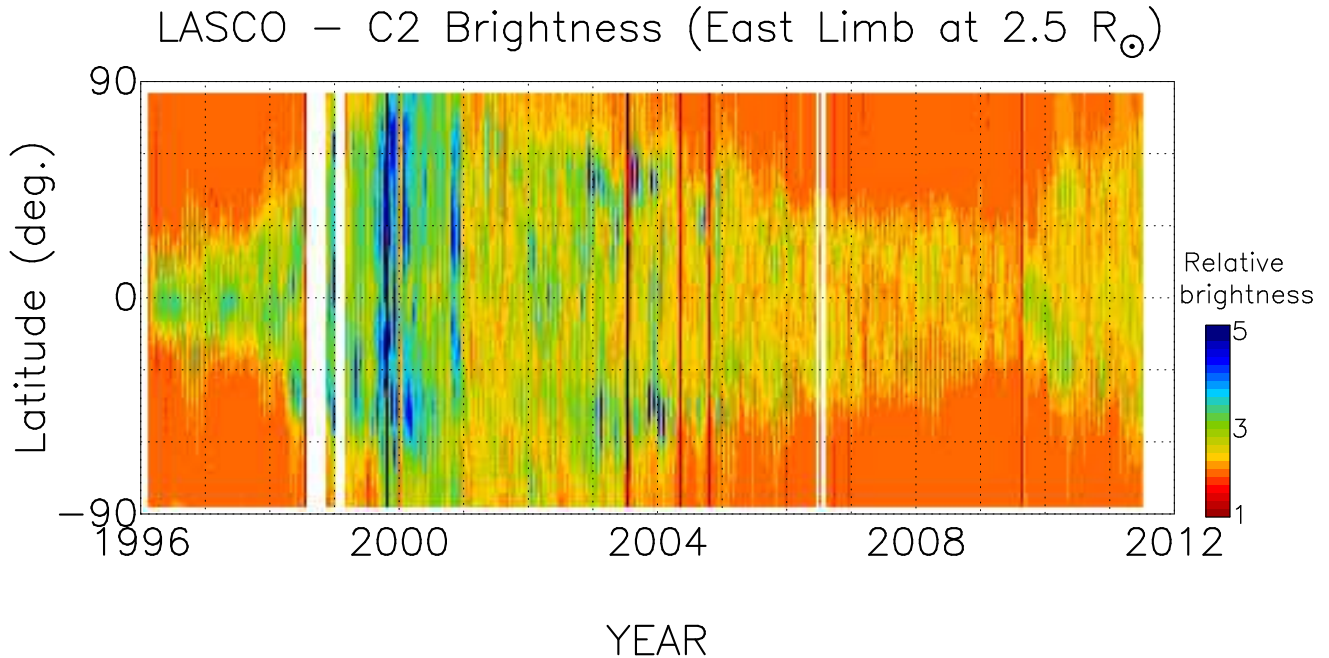}
\caption{Helio latitudinal distribution of Thomson-scattered
brightness observed at 2.5 R$_\odot$ by the LASCO-C2 coronagraph
\citep{brueckner1995}, as a function of year. The brightness
represents the coronal density. At the minimum of solar cycle 22
around the year 1996, the density above the equatorial closed-field 
corona was much intense as well as narrow in latitudinal spread, 
suggesting a nearly-dipole configuration of the coronal field. 
Whereas, in the long-decay minimum phase, the density was rather 
low along the equatorial belt and it occupied broader latitude 
range. These results are in agreement with IPS measurements.}
\end{center}
\end{figure*}

The above density turbulence image shows several interesting 
features. The high-latitude streams from coronal holes are 
always associated with low-density turbulence 
\citep{rickett1991, mano1993}. 
The drifting of large-scale structures from high to low 
latitudes, seen as intense vertical bars 
(refer to Figure 4, between years 2000 and 2004), are likely 
due to the migration of medium to large-size coronal holes 
from polar to low-latitude corona. The high-speed streams
from these coronal holes interact with the low-speed wind, causing 
the density compression and resulting in increased turbulence in 
front of the high-speed flow. The interaction is rather efficient at low 
and mid-latitudes, where the speed difference between low and 
high-speed flows is significant, but, it weakens with increasing 
latitude (e.g., Gosling et al. 1995). The spread of high-density features, 
which are limited to mid-latitude region ($\sim$$\pm$40$^\circ$) 
at the minimum phase and much wider ($>$60$^\circ$) during the 
maximum, are consistent with the tilt angle of the neutral line
(refer to Figure 7). The neutral line on the source surface of 
the Sun divides regions of opposite polarities. It is swept 
radially outward by the solar wind and forms the heliospheric 
current sheet (HCS).  At 
the time of polarity reversal, the amplitude of HCS extends to high 
latitudes, suggesting a complex closed magnetic field topology, 
resulting between the coronal holes and active regions. For example
in Figure 4, the disappearance of the solar wind \citep{lazarus2000}, 
shown by the vertical low-density structure around middle of 1999, 
is consistent with the complex as well as a rather closed-field 
corona to confine the plasma, as the Sun approaches towards the 
peak of the cycle 23 (Figure 7). 

The other interesting feature in Figure 4 is the co-rotating 
interaction regions (CIRs) dominated heliosphere resulted from 
the persistent large coronal holes in the latitude range of 
$\pm$45$^\circ$, as shown by the vertical intense density 
structure caused by the moderate compression during 2003--2004
\citep{zhang2008, abramenako2010}.
In this period, each 
solar rotation possessed more than one heliospheric current sheet, 
running almost parallel to the latitude axis, increased the 
possibility of interaction between low- and high-speed streams.
These interactions caused weak to moderate storms at the Earth
during 2003, as indicated by the geo-magnetic disturbance index, 
Ap (refer to Figure 9).

The density distribution in the maximum phases of solar cycles
22 and 23, respectively centered around years 1991 and 2001, 
shows rather remarkable differences. The low- and mid-latitude 
regions of the heliosphere near the maximum of cycle 22 were 
associated with high-density structures. Whereas the large 
portion of the heliosphere, even at the maximum of cycle 23, was 
occupied by a relatively low-density plasma, as was indicated 
by the density turbulence contours (refer to Figures 2 and 3). It is 
also interesting to note large differences in low-density 
plasma turbulence at the polar regions near the minima 
of cycles 22 and 23. During years 1996--1997, the low level of 
turbulence was confined to latitudes above 45$^\circ$ in the
north and south hemispheres.
In the prolonged minimum period, during years 2006-2009, the
density was much lower than that of the previous minimum as well as
it occupied a broader latitude range extending from high to low
latitudes. The low-density plasma prevailed 
the large portion of the inner heliosphere around the unusual deep 
minimum phase and it is in agreement with the corresponding 
shrinking of scattering diameter of the corona (Figure 3). The density
distribution observed after about the year 2010 indicates that the
heliosphere is gradually climbing towards the maximum phase of the 
solar cycle 24. The rate of increase of density to high latitude
in the north pole seems to be faster than that observed at the
south pole, suggesting that the solar cycle 24 tends to approach
a near-maximum phase in the northern hemisphere.

\begin{figure*}
\begin{center}
\includegraphics[width=8.0cm,,angle=-90]{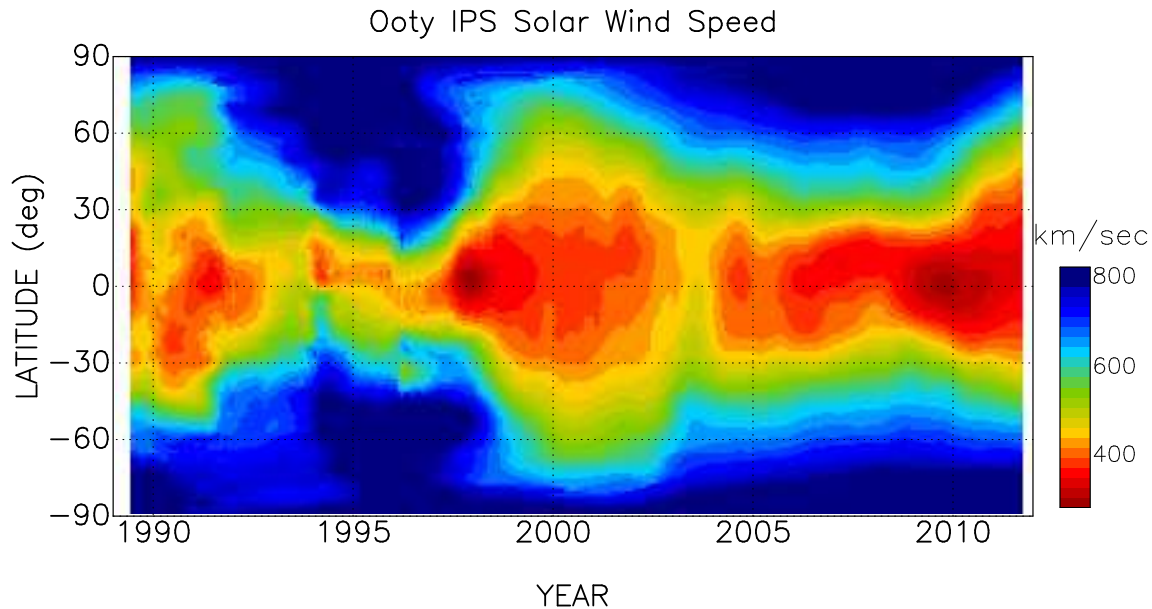}
\caption{(a) Helio latitudinal distributions of solar wind speed obtained
from Ooty IPS data plotted for years 1989--2011. This image has been made 
by tracing backward/forward from the measurement location onto a sphere of 
radius $\sim$125 solar radii. At the minimum of solar cycle 22, the effects
of high-speed wind were observed from poles up to about $\sim$30$^\circ$ 
latitudes.  But, in the prolonged minimum, the high-speed streams were 
located close to poles.}
\end{center}
\end{figure*}

\subsection{Density Distribution in the near-Sun Corona}

Ooty density turbulence results are consistent with the brightness
measured from the Thomson-scattered white-light in the near-Sun region. 
Figure 5 shows the `latitude--year' image of the white-light, which 
is associated with the density of free electrons, measured at 2.5 $R_\odot$ 
above the east limb of the Sun, for the period starting from the launch of 
LASCO coronagraph to nearly middle of the year 2011 (refer to Brueckner 
et al.~(1995) for details on LASCO coronagraph). As shown by the 
IPS data, the heliosphere was prevailed by low density during the 
prolonged minimum of the solar cycle 23. The latitudinal extents of 
large-scale low density structures in the mid- and high-latitude 
regions are also agreement with the IPS density turbulence 
distributions. The intense density structure observed
on the IPS image, associated with CIRs generated by the mid- and
low-latitude coronal holes around 2003, is represented on the LASCO
image by a less-brightness stripe, which provides evidence for the
low-density wind originating above the coronal hole and in particular,
not yet developed interaction as well as compression at the closer
solar offset. 
In the minimum phase of cycle 23, the long-lived coronal holes observed 
during 2003, might have a more direct involvement in shaping the magnetic 
state of corona at the deep minimum phase \citep{abramenako2010}.
The three-dimensional distribution of Lyman-$\alpha$ brightness, which
reflects the ionization of the neutral gas by the solar wind, also in
agreement and showed a latitudinal distribution similar to the above 
shown density turbulence and density images, respectively, displayed in 
Figures 4 and 5 (Lallement et al. 2010). The onset of the solar cycle 24
in the year 2009 and the gradual increase in the latitudinal 
extent of high density wind to higher latitude between 2010 and 2011 
(particularly at the north-pole side) are evident in Figure 5. However, 
as indicated by the density turbulence (Figure 4), the southern 
hemisphere seems to progress slower than the northern part.

\section{Solar Cycle Changes of Solar Wind Speed}

For an IPS observation in the weak-scattering regime, a suitable calibration 
of the temporal power spectrum of intensity fluctuations, having sufficient 
signal-to-noise ratio (i.e., $\gtrsim$15 dB), can provide the speed of the solar wind 
\citep{mano1990, mano2000, tokumaru1994, 
tokumaru2011, yamauchi1996, yamauchi1998, liu2010}.
Figure 6 shows the `latitude--year' distribution of solar wind speed estimates 
obtained from the IPS data collected  from the Ooty Radio Telescope, as it 
was displayed in the plots of density and density turbulence (Figures 4 and 5). 
During the minimum of solar cycle 22, polar regions were dominated by high speed 
streams in the range $\sim$700--800 km~s$^{-1}$, from open-field coronal holes 
and low and variable flow speeds, $\leq$450 km~s$^{-1}$, were observed at the 
low- and mid-latitude regions of the complex/closed field corona 
\citep{phillips1995, mccomas2008, smith2008}.
The striking features in this image are (i) confined period of minimum for 
cycle 22, centered around year 1996 and the high-speed wind (or the extension 
of coronal holes) observed from poles to mid-latitude region of $\sim\pm30^\circ$, 
and (ii) in contrast, the effects of minimum of cycle 23 was stretched over 
a long-period of time and the width of the high-speed belt was limited to
latitudes higher than 60$^\circ$ in north and south poles; a deep minimum-like 
condition was observed between years 2008 and 2009, during when the extent of 
high-speed belts at the north and south poles were limited between the poles 
and $\sim$60$^\circ$ latitudes. Further, the speed originating above them has 
been considerably reduced. The low-speed solar wind distribution, along the equatorial 
belt, seems to be highly variable in the range 300--500 km~s$^{-1}$, and the 
`disappearance of solar wind' (i.e., extremely low density and speed)
occurred in the early part of the year 1999 as well 
as the coronal-hole dominance in the mid-latitude region of the heliosphere during 
2003 are evidently revealed. 

Figure 7 shows the smoothed plots of tilt angle of the magnetic heliospheric 
current sheet (HCS), and strength of polar magnetic field. For reference,  
sunspot number and radio flux density at 10.7 cm are also included in this 
figure. The magnetic field data sets have been obtained from 
the Wilcox Solar Observatory\footnote{\url{http://wso.stanford.edu}}
and the solar activity indexes are from Solar Geophysical Data 
Center\footnote{\url{http://www.ngdc.noaa.gov/stp}}.
These plots cover a period from 1985 to the later 
part of 2011, i.e., solar cycles 22 to 24. The large-scale magnetic field, shown 
by the amplitude of current sheet during years 2004--2008, resembles a 
corona of complex field, but with the reduced activity. The remarkable increase 
in the latitudinal width of low-speed flow during the extended minimum of cycle 
23 is directly linked and in good agreement with the observed polar field strength 
and warping of the current sheet, which is caused by the extension of global field 
from the Sun into the interplanetary medium. The solar wind speed distributions 
observed for solar cycles 22 and 23 correlate, respectively, with the polar field 
strength of $\sim$100 to 50$\mu T$. Thus, the magnetic field has been  weak by 
$\sim$40--50\% in the minimum phase of cycle 23 than that of cycle 22 (Figure 7). 
The magnetic pressure associated with the polar coronal holes consequently seems 
to play a significant role in the 
acceleration of high-speed wind. The weak field may be due to the fact that 
the polar field has not fully developed after the field reversal between 2000 
and 2002 (Figure 7). 

\begin{figure}
\begin{center}
\includegraphics[width=8.0cm]{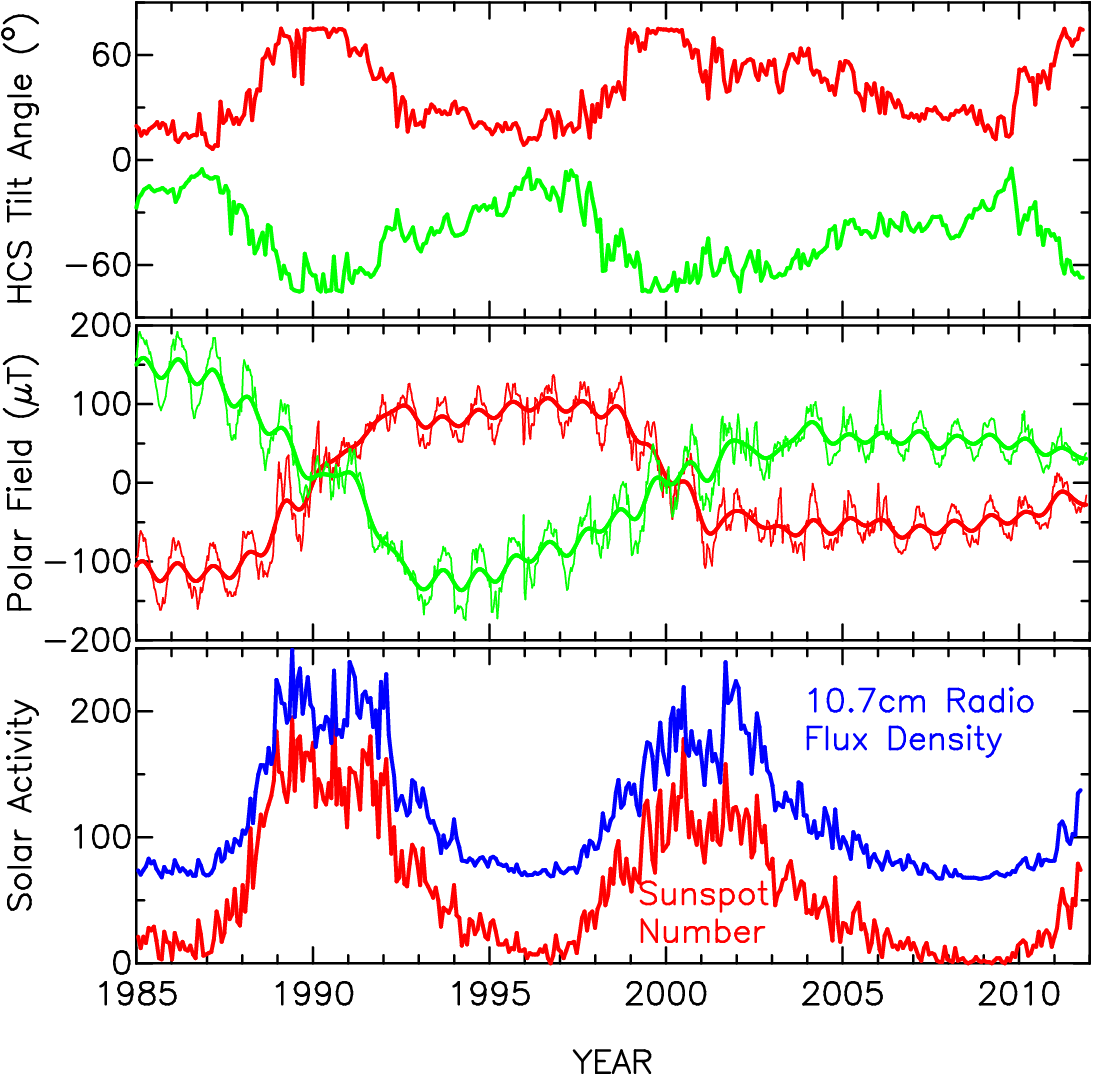}
\caption{The stack plot shows solar activities (i.e., sunspot number and 
solar radio flux density at 10.7 cm), intensity of polar magnetic field, 
and tilt angle of heliospheric current sheet (HCS). The reduction in the 
strength of the polar fields is evident after the reversal of the field
around the year 2000.}
\end{center}
\end{figure}

It was reported that the Sun went through a period of large number of 
`sunspot-free' days\footnote{\url{http://sidc.oma.be/sunspot-data/}}
(more than 800 days between 2006 and 2009, in 
comparison with $\sim$300 days of `spotless' days in the minimum of cycle
22). Moreover, at the extreme minimum phase of cycle 23, the solar wind 
distribution around the equatorial belt as well as the tilt angle (refer 
to Figure 7) suggest that the magnetic field of the Sun did not approach 
the expected dipole geometry, as it did for the minimum  phase of cycle 22 
\citep{riley2003, tokumaru2009}.
The weakening of the large-scale coronal field has also been revealed by 
the weak emission of Fe XIV green line at $\lambda$5303 \AA~
\citep{rybansk2005}.
Additionally, Ulysses and near-Earth observations (Figures 8 and 9) made over the solar 
cycles 22 and 23 are in good agreement with speed and density distributions
presented in the previous sections \citep{mccomas2008, smith2008, lee2009}.
Figure 8 shows the daily averages of the magnitude of interplanetary 
magnetic field and solar wind speed, as measured by Ulysses spacecraft
around minima of solar cycles 22 and 23. Ulysses measurements
evidently show an overall reduction in field strength, density, speed, 
as well as the increased width of the equatorial flow and poleward 
shrinking of the high-speed wind for the similar latitudinal and 
heliodistance passes of cycles 22 and 23.  
The lack of sunspots activity 
governing the radiative energy from Sun, in combination with the 
weakening of the interplanetary field and turbulence, allowed the 
penetration of cosmic rays at the minimum phase of cycle 23 by more 
than $\sim$20\% than at 1997--1998 period (Mewaldt el al. 2010).

\begin{figure}
\begin{center}
\includegraphics[width=8.0cm]{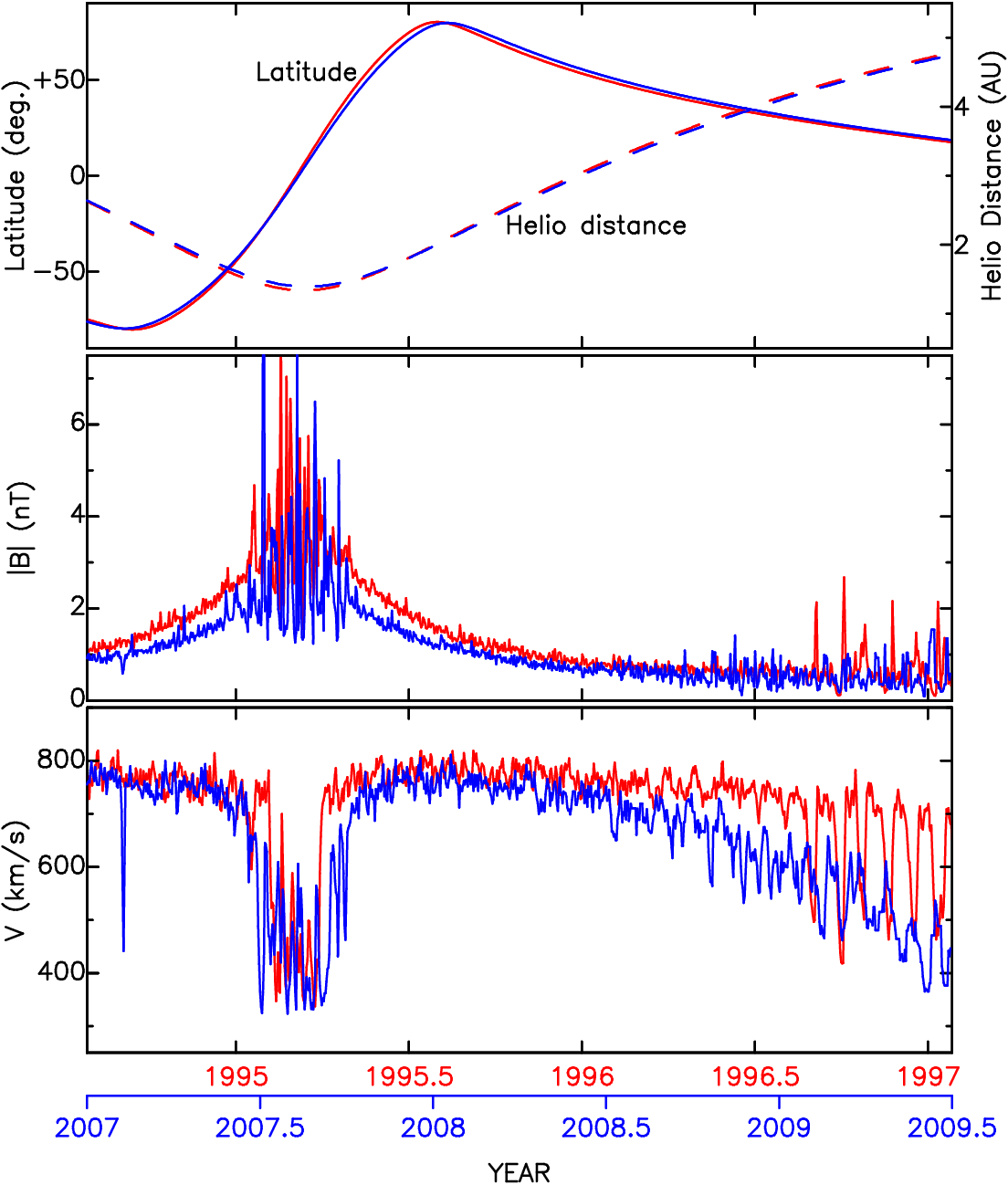}
\caption{Daily averages of the magnitude of interplanetary magnetic
field and solar wind speed at the orbit of Ulysses as a function of
year, around minimum phases of solar cycles 22 and 23. Close tracking 
of location of Ulysses on the heliosphere, for these two passes, clearly
shows (i) significant reduction in equatorial flow widths, (ii) rapid
decline in speed, and weaker field at minimum of cycle 23 than that of
corresponding previous minimum of cycle 22.}
\end{center}
\end{figure}

\begin{figure*}
\begin{center}
\includegraphics[width=12.7cm]{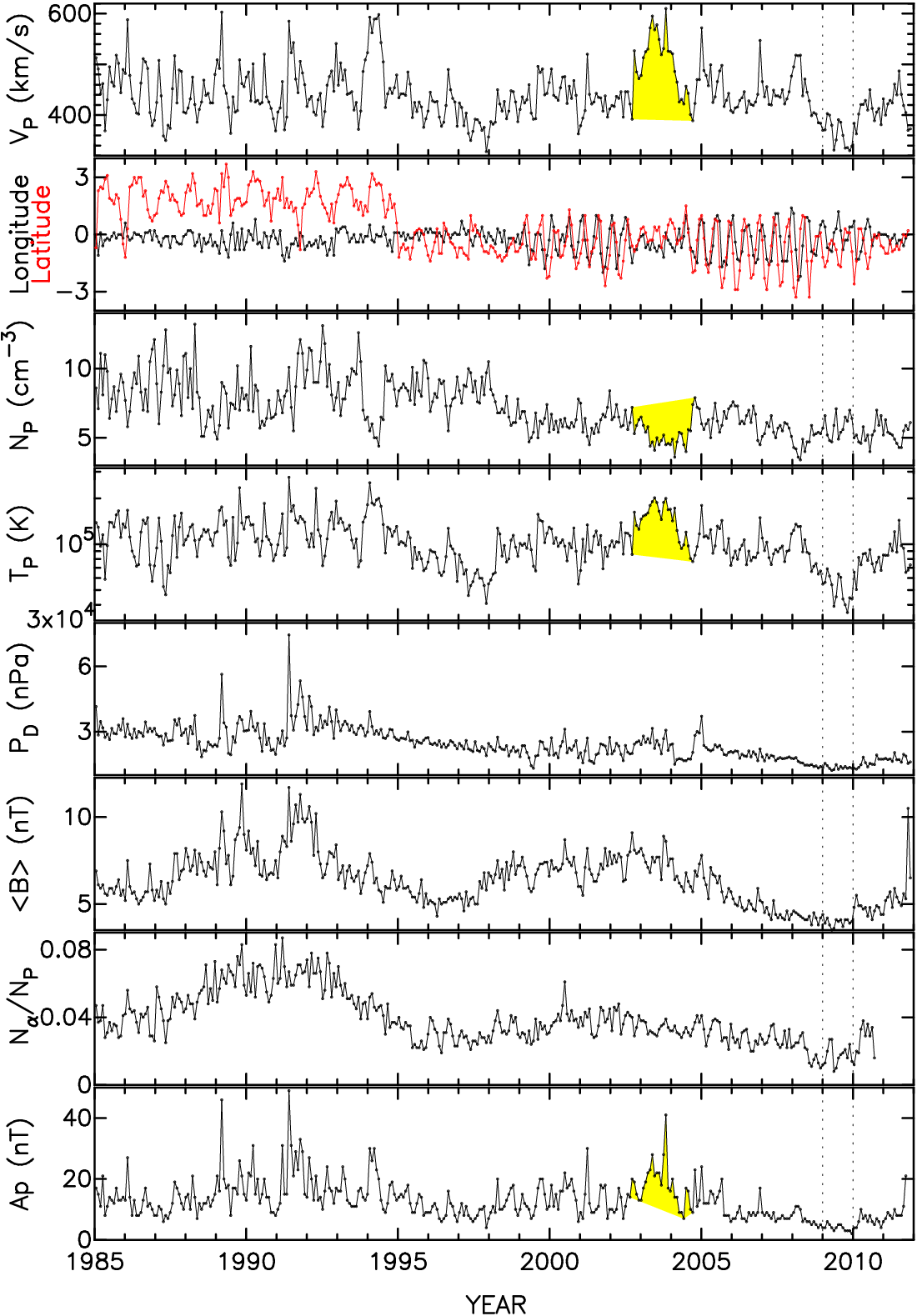}
\caption{Plots of 27-day average of solar wind proton speed and flow 
directions (i.e., latitude and longitude in degree), number density, 
temperature, dynamical flow pressure, magnitude of interplanetary 
magnetic field, and ratio of alpha particle to proton densities, 
as a function of year. The geo-magnetic index, Ap, is shown in the
bottom plot. The yellow shades show CIR dominated period during 
the minimum of cycle 23. The vertical dotted lines indicates the 
deep minimum phase of cycle 23.}
\end{center}
\end{figure*}

As observed in density and density turbulence plots (refer to Figures
4 and 5), the speed distribution plot (Figure 6) also shows that after the 
onset (i.e., in the ascending phase) of solar cycle 24, starting from 
middle of 2009, the changes observed in the distribution of high-speed 
solar wind at the north and south polar regions are different. In 
particular, at the later part year 2011, the high-speed belt at the 
northern region has moved almost close to the its pole. However, in 
the southern part, it has remained nearly unchanged. It suggests that 
the area of the northern coronal hole associated with the high speed 
wind has reduced and shrunk close to the north pole. In other words, 
the increase in the solar activity (i.e., vanishing of source of
high-speed wind) is significant at the northern 
hemisphere and indicates nearly (or close to) the maximum phase of 
cycle 24. It is consistent with the yearly average values of flare 
index\footnote{\url{ftp://ftp.ngdc.noaa.gov/STP/SOLAR\_DATA}}
observed at northern and southern hemispheres over the year 2010, 
respectively, 0.26 and 0.12. The northern hemisphere of the Sun 
was 2 times more active than the southern hemisphere.

\section{Solar Cycle Changes at near-Earth Space}

Figure 9 shows the 27-day average of 1--AU measurements of solar wind 
proton speed and its flow directions (i.e., latitude and longitude), 
number density, 
temperature, dynamical flow pressure, magnitude of interplanetary 
magnetic field, and ratio of alpha particle to proton densities, 
observed between years 1985 and 2011. A plot of geo-magnetic index, 
Ap, is also included in the figure. These data sets have been 
extracted from the NASA/GSFC's OMNI data 
base\footnote{\url{http://omniweb.gsfc.nasa.gov/}}.
It is evident that in the near-Earth 
region, the solar wind dynamical pressure started to decrease after 
the maximum phase of cycle 22 (after about year 1992), to a deep 
minimum during the years 2008--2009. Although there was a moderate 
enhancement in the solar wind pressure at the maximum of cycle 23, 
it was however nearly 40\% lower than the corresponding increase 
observed at the previous activity maximum. In the long decay phase
of cycle 23, the inter comparison of results from previous sections
with 1--AU measurements of solar wind speed, proton and 
alpha particle densities, temperature, pressure and interplanetary
magnetic field suggests that the reduction in density and magnetic 
energy of the solar wind streams has organized the heliosphere. 
Thus, the energy of the solar wind started to decrease after the 
maximum of the cycle 22 and continued up to the transition between
cycles 23 and 24.

In Figure 9, there are two interesting periods, 1993--1994 and 2003--2004, 
in which the presence of equatorial coronal holes dominated the solar
wind flow.  They are prominently seen as enhancement in speed and 
temperature, and depletion in density. The period, 2003-2004, is shown 
in shade on plots of speed, density, temperature and corresponding 
geo-magnetic index. As revealed in the speed and density 
turbulence images (Figures 4 and 6), the corresponding 
interaction between the high-speed wind from low-latitude coronal hole
and ambient (or background) low-speed wind has resulted in compression 
region and led to increase in density turbulence (refer to Figures 4 and 6).
The period of coronal hole dominated low-latitude solar wind was shorter 
during the cycle 22 than that of cycle 23, in which the solar wind interactions 
have caused moderate to strong geo-magnetic activities as recorded by
Ap index. 

The presence of equatorial and mid-latitude coronal holes and their
evolution during the years 2003--2004 have likely caused the changes
observed in the flow direction of the solar wind. In Figure 9, the 
second plot from the top shows
the flow longitude and latitude. The solar wind longitude, during years 
1985 to 1999, did not show systematic variation with the activity of 
the Sun. Whereas the quasi-periodic oscillations of amplitude 
$\sim$2--4$^\circ$ observed in the solar
wind flow latitude were likely caused by the latitudinal warping of the 
heliospheric current sheet. In contrast, the regular variations in flow
latitude and longitude ($\sim$3--4$^\circ$) between 1999 and 2003 show
that the flow direction exhibits cyclic changes in the west-north-east-south 
direction. The mean period of oscillation is $\sim$6 month, which indicates
a cyclic behavior linked to the slow changes ($>$27-day period)
of large-scale structures of the source of the solar wind. However, the 
above cyclic pattern of solar wind flow direction disappears between 2003 
and middle of 2004, in coincidence with the duration of the high-speed wind 
caused by the large mid-latitude coronal hole. But, it appears back in the later 
subsequent time and the solar wind flow geometry changes to a reversed 
opposite cyclic behavior of east-north-west-south pattern. 

In the ascending phase of solar cycle 23, after the year 1994, the 
dominance of equatorial coronal holes disappeared, the solar wind 
flow direction switched from largely north pointing to equatorial flow.
It implies that the large-scale magnetic field averaged over a solar
rotation, with respect to the heliospheric equator, has gone through a 
systematic reversal during the ascending phase of the solar cycle. 
In the case of solar cycle 23, the persistence of the cyclic pattern from 
2004 to 2011 indicates the continued presence of large-scale magnetic 
configuration of complex field, which may be caused by the group of 
long-lived small coronal holes of opposite polarities in the low-latitude 
region of the Sun.
Thus, the net effect of formation and decay of coronal
holes, as a result of magnetic reconnection between global and active
region fields \citep{fox1998},
essentially has determined the magnetic field configuration of the corona
and heliosphere as well as the rate of occurrence of CIRs and associated
transients \citep{zhang2008, yashiro2005}.

\section{Discussion and Summary}

This paper presents the analysis of three-dimensional evolution of 
large-scale structures of
solar wind density turbulence and speed at different levels 
of solar activity between 1985 and 2011, which includes solar cycles 22
to 24. The long-term IPS observations obtained from the Ooty Radio 
Telescope at 327 MHz, supplemented with other ground- and space-based 
measurements, show that the solar cycle changes in the solar wind are
significantly different between cycles 22 and 24. The main results of 
this study are as follow:

(1) The value of density turbulence in the inner-heliospheric region 
was the highest at the maximum of solar cycle 22 (around the year
1991) and decreased to a level of $\sim$70\% in the subsequent minimum 
phase (around the 1996). A similar reduced trend was also observed at 
the transition phase between cycles 21 and 22 (around the year 1986).

(2) However, at the deepest minimum of the cycle 23 at 2009, the 
density turbulence decreased to a level of $\sim$30\% in comparison to 
the highest level observed at the maximum of the cycle 22 at 1991. It 
indicates that during the long decay phase of cycle 23, the source 
region of the solar wind on the Sun has experienced severe deficiency 
in energy as well as density.

(3) The important result of this study is that the scattering diameter
of the corona (i.e., density turbulence contours displayed in Figure 3) 
has steadily decreased after about 2003 and attained the smallest size 
during middle of 2009. For a typical radial fall of scattering power, 
$R^{-4}$ to $R^{-4.4}$, 
it suggests a reduction of more than 60\%. The gradual decrease in the 
scattering power of the corona is consistent with the globular 
downward changes observed in the strength of solar magnetic field, 
leading to a reduction in the supply of mass and energy at the base of 
the corona and into the heliosphere.

(4) The three-dimensional results of solar wind speed also show 
remarkable changes in the latitudinal distributions of high- and 
low-speed flows between solar cycle 22 and 23. For example, the 
source region of high-speed wind (i.e., $\sim$700-800 km~s$^{-1}$)
at the minimum of cycle 22 was wide in latitude and extended from 
poles to mid latitudes of $\sim$30$^\circ$. However, during 
2006--2009, the high-speed regions were narrow in latitude and 
confined close to the poles. Thus, in the long-decay phase of cycle 
23, the heliosphere encountered a net decrease in solar wind speed at 
most of the latitudes.

(5) Both speed and density turbulence distributions obtained from
IPS are consistent not only with the reduction of solar activity 
but, also relatively complex corona for most of the minimum phase
of the cycle 23. Results on the latitudinal spread suggest 
that the solar corona did not reach the simple ``dipole" shape often 
observed during solar minima, while low-latitude coronal holes and their 
associated co-rotating high-speed solar wind streams persisted until 
the deepest minimum and caused large amplitude HCS, which heavily 
modulated the solar wind and opposed the formation of dipole-shaped 
corona.

(6) The results from IPS and LASCO confirm the onset and growth of the 
solar cycle 24, starting from about middle of 2009. It is interesting 
that the high-speed wind (also high-density plasma) at the northern 
side has almost moved close to the pole, indicating a reduced area of 
the coronal hole at a phase similar to nearly approaching maximum phase 
of the cycle. But, in the southern hemisphere, the activity has yet to 
develop. The question that how far the maximum of the current cycle 
will rise.

In the decay phase of cycle 23, the reasons for the reduction in 
global field strength, density, and 
flow speed are possibly due to changes in the movement of large-scale 
fields, as the reversal of polarity progresses. It corresponds to the 
rate of poleward and equator-ward meridional flows, which act as the 
conveyor belt in transporting magnetic flux (i.e., plasma and frozen 
in magnetic field) at the solar interior. In the deep minimum phase, 
the weak fields observed at the poles are likely to be associated with 
the transport of unbalanced flux by the meridional flow and a faster 
flow rate (relative to diffusion) will result in less unbalanced flux 
in each hemisphere as well as causes weaker fields at the poles 
\citep[e.g.,][]{sheeley2010}.

\acknowledgments 
All the members of the Radio Astronomy Centre (NCRA-TIFR) are 
acknowledged for making the Ooty Radio Telescope available for 
IPS observations. The sunspot number and the radio flux density 
at 10.7 cm were obtained from the National Geophysical Data Center. 
The international Sunspot Number data were provided by the Solar 
Influences Data Center (\url{http://sidc.oma.be/sunspot-data/}). 
The near-Earth solar wind data and geo-magnetic indexes were 
obtained from OMNIWeb service and OMNIdata of NASA/GSFC's Space 
Physics Data Facility. Ulysses data sets were provided by 
Dr. A. Balogh (magnetic field) and Dr. J.L. Philips (plasma) 
via CDAWeb, maintained at Space Physics Data Facility, NASA/GSFC.
The Wilcox Solar Observatory provided the source surface magnetic field 
data. SOHO (LASCO) is a project of international cooperation between 
ESA and NASA.  This work was partially supported by the CAWSES-India 
Program, sponsored by ISRO.

\end{document}